\documentclass[a4paper]{jpconf}
\usepackage{graphicx}
\usepackage{cite}

\newcommand{\gtrsim}{ \mathop{}_{\textstyle \sim}^{\textstyle >} }

\begin{document}
\title{Neutrino mass constraint from CMB and its degeneracy with other cosmological parameters}

\author{Kazuhide Ichikawa}

\address{Institute for Cosmic Ray Research, University of Tokyo, Kashiwa 277-8582, Japan}

\ead{kazuhide@icrr.u-tokyo.ac.jp}

\begin{abstract}
We show that the cosmic microwave background (CMB) data of WMAP can give subelectronvolt limit on the neutrino mass: $m_\nu < 0.63$\,eV (95\% CL). We also investigate its degeneracy with other cosmological parameters. In particular, we show the Hubble constant derived from the WMAP data decreases considerably when the neutrino mass is a few times 0.1\,eV.
\end{abstract}

\section{Introduction} \label{sec:introduction}

It is known that cosmological considerations give more stringent constraints on the neutrino mass than the present tritium $\beta$-decay experiments.
For example, \cite{Spergel:2003cb} obtained $m_\nu < 0.21$\,eV and \cite{Tegmark:2003ud} obtained $m_\nu < 0.58$\,eV, using the WMAP 1st year data (WMAP1) combined with the galaxy power spectrum (the former used the data from the 2dFGRS and the latter from the SDSS main sample. The difference can be ascribed to the use of the bias information by \cite{Spergel:2003cb}). Meanwhile, we found $m_\nu < 0.66$\,eV from WMAP1 alone as reported in \cite{Ichikawa:2004zi}. We note that this is the first to point out CMB data (precisely speaking, WMAP1) alone can give a sub-eV upper bound on the neutrino mass, which is comparable to the limits obtained from the CMB and galaxy clustering data combined. 

Before \cite{Ichikawa:2004zi} (and some time after that too), there seems to be a lack of consensus about whether the CMB experiment with the WMAP-level precision can derive a sub-eV neutrino mass limit and, in fact, the WMAP1 alone limit reported in \cite{Tegmark:2003ud}, $m_\nu < 3.8$\,eV, which allows 100\% HDM was seemingly accepted (incidentally, the WMAP group did not report the WMAP1 alone limit on the neutrino mass). We, on the contrary, have derived the upper limit 0.66\,eV as quoted above from the same data by the $\chi^2$ minimization method which is independent from the MCMC method adopted by \cite{Tegmark:2003ud}. Our conclusion is later confirmed by \cite{MacTavish:2005yk,Hannestad:2006zg,Lesgourgues:2006nd} (\cite{MacTavish:2005yk} does not report the WMAP1 alone limit in a number but judging from their likelihood figure, it looks less than 1\,eV. \cite{Hannestad:2006zg} and \cite{Lesgourgues:2006nd} gives $m_\nu < 0.70$\,eV and $m_\nu < 0.63$\,eV respectively).

Below, we first discuss the WMAP alone limit on the neutrino mass comparing results from WMAP1 and the WMAP 3rd year data (WMAP3). Then, we will investigate its degeneracy with other cosmological parameters, especially the Hubble constant. 

We here summarize our notations. We derive neutrino mass constraint in the flat $\Lambda$CDM
model with the power-law adiabatic perturbations. Namely, cosmological parameters we consider are:
 baryon density $\omega_b$, matter 
density $\omega_m$, hubble parameter $h$, reionization optical depth $\tau$,
spectral index of primordial spectrum $n_s$, its amplitude $A$ and 
massive neutrino density $\omega_\nu$. Here, $\omega \equiv \Omega h^2$ 
where $\Omega$ is the energy density normalized to the critical density and
$\omega_m \equiv \omega_b + \omega_c$ where $\omega_c$ is CDM density
(Caution that many literatures define $\omega_m \equiv \omega_b + \omega_c +\omega_\nu$ to include the massive neutrino in the matter density).
$\omega_\nu$ is related to neutrino masses by $\omega_\nu =
\sum m_\nu/(94.1$\,eV) and we assume three mass degenerate generations of neutrinos so that $\omega_\nu =m_\nu/(31.4$\,eV). 

\section{WMAP alone limit} \label{sec:limit_WMAP}

\begin{figure}
\begin{minipage}{7.5cm}
\scalebox{0.55}{\includegraphics{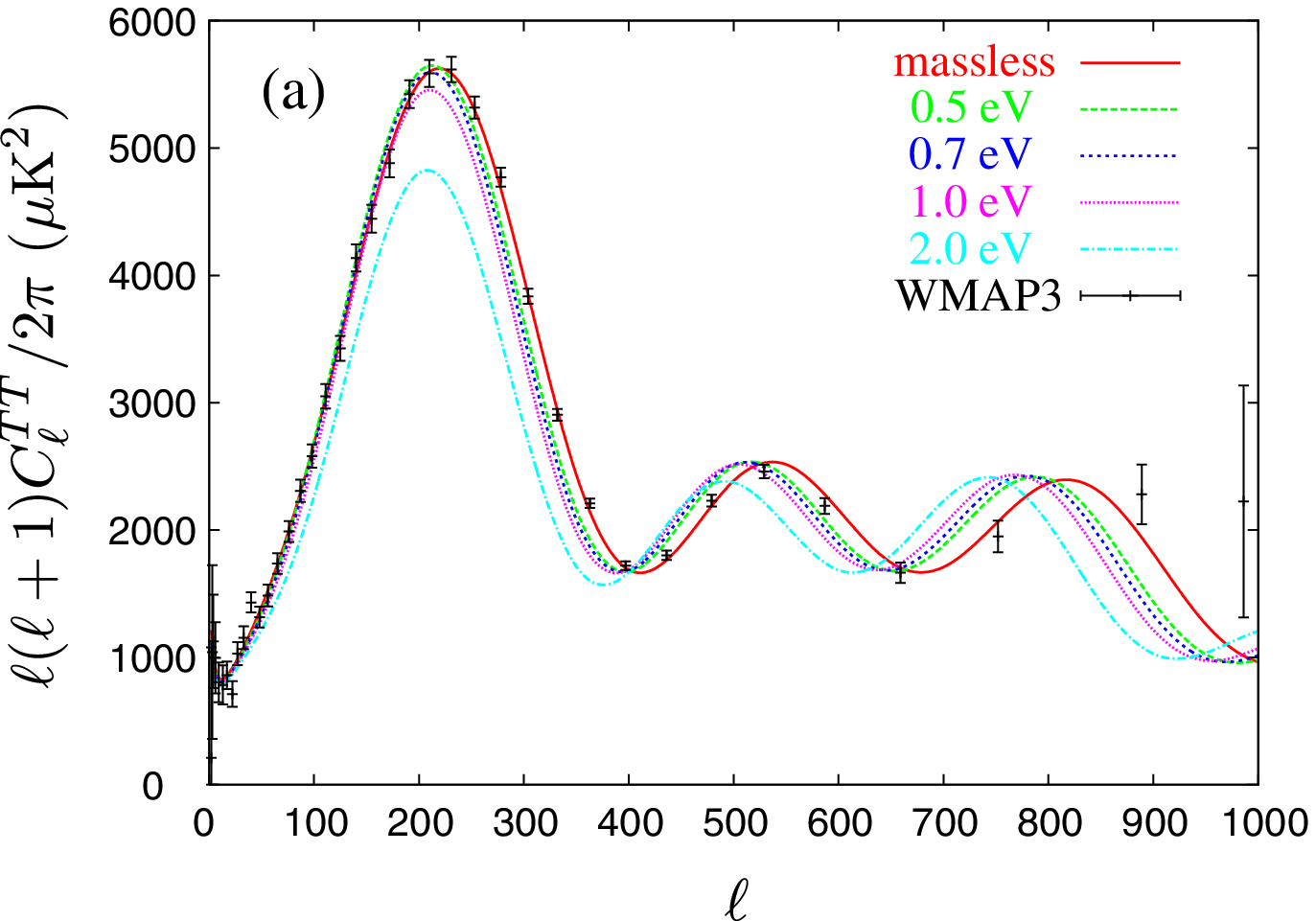}}
\end{minipage}
\begin{minipage}{7.5cm}
\vspace{-0.28cm}
\scalebox{0.603}{\includegraphics{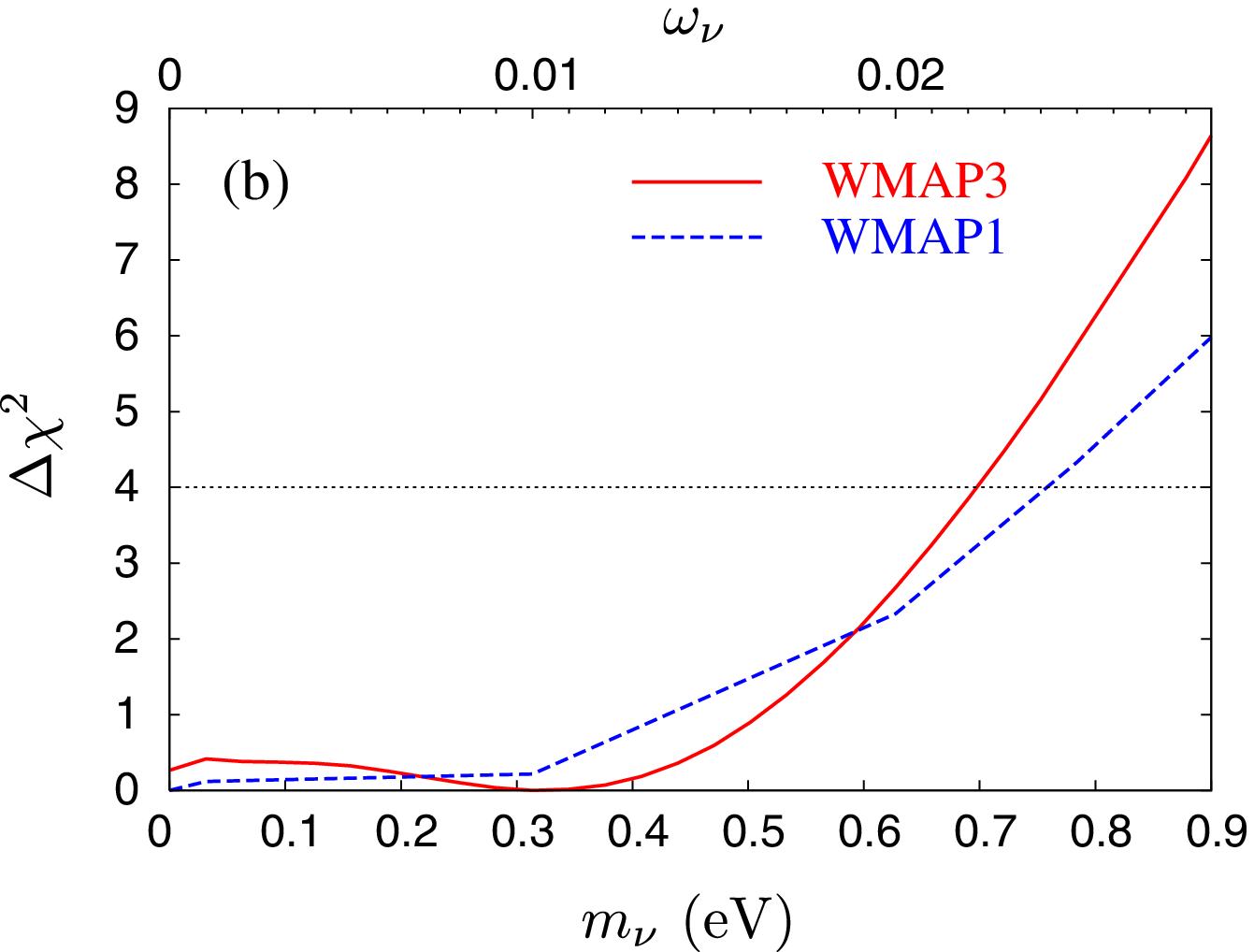}}
\end{minipage}
\caption{(a) Effects of massive neutrinos on the CMB TT power spectrum. 
(b) $\Delta \chi^2$ of WMAP data as functions of the neutrino mass $m_\nu$. $\Delta \chi^2=4$ roughly corresponds to 95\% C.L. limit. 
The blue dashed line uses the WMAP1 (TT+TE) \cite{Ichikawa:2004zi} and the red solid line uses the full WMAP3 including temperature and polarization.}
\label{fig:CMB_numass}
\end{figure}

We begin with showing how CMB power spectrum is modified by increasing neutrino mass in Fig.~\ref{fig:CMB_numass} (a).
The other cosmological parameters are fixed here. 
In this figure, we see horizontal shift and suppression around the first peak. 
The horizontal shift comes from the fact that the larger $m_\nu$ (more non-relativistic particles at present epoch) implies that the distance to the last scattering surface is shorter and the peaks move to smaller $\ell$. 
However, this shift is easily cancelled by the shift in $h$. 
Therefore this does not produce a neutrino mass signal. 
The suppression of the 1st peak takes place only when $m_\nu \gtrsim 0.6$\,eV.
This corresponds to 0.3\,eV in terms of photon temperature $T_\gamma$. Meanwhile, the recombination takes place at $z \approx 1088$ or $T_\gamma \approx 0.3$\,eV. In other words, massive neutrinos become non-relativistic before the epoch of recombination 
if they are heavier than 0.6\,eV. Therefore, only in this case, the neutrino mass can significantly imprint a characteristic signal in acoustic peaks (specifically, the matter-radiation equality occurs earlier due to less relativistic degrees of freedom and the enhancement of the 1st peak by the early-integrated Sachs-Wolfe effect is smaller). 

This signal, however, could be accidentally mimicked by some combination of other cosmological parameters.
 So we searched a large cosmological parameter space in order to find the degree of degeneracy between $m_\nu$ and the other cosmological parameters. 
For each value of $m_\nu$, we varied 6 other $\Lambda$CDM cosmological parameters to find minimum $\chi^2$. 
The results are shown in Fig.~\ref{fig:CMB_numass} (b) \cite{Fukugita:2006rm}. The WMAP1 result \cite{Ichikawa:2004zi} is also shown. 
We obtained the upper bound of 0.63\,eV from WMAP3 alone\footnote{
There is small difference from the limit reported in \cite{Fukugita:2006rm} (0.68\,eV) because we here report the one obtained with the updated likelihood code (ver.~Nov.~2006) by the WMAP team. Our new limit here is consistent with that of published version of the WMAP 3-year paper \cite{Spergel:2006hy}, $m_\nu < 0.60$\,eV.
}
and notice that
the WMAP3 constraint is not improved much from the WMAP1 limit.
This is reasonable because the neutrino mass (larger than 0.6\,eV) characteristically modifies the acoustic peaks
around 1st and 2nd peaks in the temperature power spectrum and these regions are already well measured 
by the WMAP1.

We stress that this bound is robust in a sense that
it is obtained from CMB data of the WMAP which is considered to be the cleanest cosmological data and
that it is obtained from a single experiment.
Also, CMB can be dealt with the linear perturbation theory so it does not suffer from non-linearity or biasing
which appear in galaxy clustering data.

\section{The degeneracy between the neutrino mass and other cosmological parameters} \label{sec:degeneracy}
\begin{figure}
\begin{center}
\scalebox{0.6}{\includegraphics{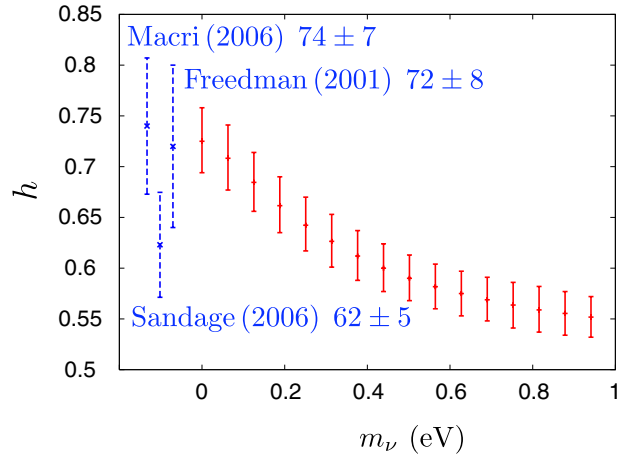}}
\caption{The constraints on $h$ for several fixed values of neutrino mass (red solid bars). The constraints from distance ladder measurements \cite{Freedman:2000cf,Sandage:2006cv,Macri:2006wm} are also shown (blue dashed bars).}
\end{center}
\label{fig:on_h}
\end{figure}

So far, we have argued that the Hubble constant is degenerate with the neutrino mass in the direction to keep the acoustic peak position constant. Namely, when $m_\nu$ is larger, $h$ should be smaller in order to fit the CMB data. We show this explicitly in 
Fig.~2 for the case with WMAP3. Note that if $m_\nu \gtrsim 0.3$\,eV, the best fit $h$ is significantly lower than $h = 0.7$, which is usually considered as the WMAP value assuming massless neutrinos. More interestingly, among the measurements of $h$ via the cosmic distance ladder, there is a group \cite{Sandage:2006cv} deriving somewhat lower $h$ than the famous value of \cite{Freedman:2000cf}. We can expect both ``direct" measurement of $h$ and $m_\nu$ will improve in near future: \cite{Macri:2006wm} notes that 1\% accuracy could be obtained through measuring maser distance to a large number of galaxies in the Hubble flow, and \cite{Angrik:2005ep}
states that $m_\nu = 0.35$\,eV could be determined with 5$\sigma$ by KATRIN experiment. I would conclude that uncertainty of $m_\nu$ is one of the largest systematic errors for estimating cosmological parameters from CMB. We have here shown the effect on $h$ but other parameters are affected too (see Fig.~2 in \cite{Ichikawa:2004zi}). In particular, the negative correlation between $n_s$ and $m_\nu$ is notable.

\end{document}